\documentclass[a4paper,11pt]{article}
\usepackage{pos}
\usepackage[greek,british]{babel}
\def\AnswerYes{y}
\def\ShowLineNumberVersion{n}     %%%%%% Version with line numbers or final
\def\ShowLabelsVersion{n}         %%%%%% Show defs. of refs, cites or final
\def\ShowChangesVersion{n}        %%%%%% Changes highlighted or final version
\def\ShowAnnotationsVersion{n}    %%%%%% Version with annotations or final
\def\feynVersion{n}               %%%%%% Choose whether feynman graphs to be
\usepackage{color}
   \usepackage{xparse}
   \NewDocumentCommand{\arxiv} %
   {r [: u{ [} u{]]} }{[\href{http://arxiv.org/abs/#2}{arXiv:#2}~[#3]]}
   \NewDocumentCommand{\arxivold} {r[]}{[\href{http://arxiv.org/abs/#1}{#1}]}
   \NewDocumentCommand{\arXiv} %
   {r [: u{ [} u{]]} }{[\href{http://arxiv.org/abs/#2}{arXiv:#2}~[#3]]}
   \NewDocumentCommand{\arXivold} {r[]}{[\href{http://arxiv.org/abs/#1}{#1}]}
\usepackage{doi} % \doi{} converts DOI numbers into clickable links
\usepackage{orcidlink}
\usepackage[UKenglish]{isodate} % correct (British) date format
\usepackage[numbers,sort&compress]{natbib}%%% collapses [1,2,3] -> [1-3] etc
\usepackage{graphicx}          %%% for pictures and figures
\graphicspath{{figures/}{./}}
\usepackage{multirow}                   %%% tabs over multiple rows
\usepackage{amsmath}
\usepackage{marvosym} % \Lightning
\usepackage{bbm} % double-line symbols (RR), with numerals (11) and lower-case
\usepackage{textcomp}% for \textinterrobang and others
\usepackage{xspace}                    %%% correct spacing self-defined macros
\xspaceaddexceptions{[]}
\usepackage{dcolumn}                    %%% Align table columns on decimal point
\usepackage{pbox}

\usepackage{bm}                        %%% bold math

\usepackage{cancel}	                  % produce backslash though symbol

\usepackage{rotating}

\usepackage{floatpag}           % eliminate page number from pages with
\floatpagestyle{plain}          % ordinary behaviour
\usepackage{slashed}
\ifx\feynVersion\AnswerYes
   \usepackage{feynmp-auto} % for Feynman diagrams; nice definitions after \begin{document}
\fi
\usepackage{chngcntr} % allows one to reset counters between sections etc
\counterwithin*{equation}{section} % reset equation counter with each section

\ifx\ShowLabelsVersion\AnswerYes
   \usepackage[color]{showkeys} 
   \definecolor{refkey}{gray}{.3}%{.5}   % slightly gray font
   \definecolor{labelkey}{gray}{.3}%{.5} % slightly gray font
   
\fi
\ifx\ShowAnnotationsVersion\AnswerYes
   \newcommand{\comment}[1]{{\scriptsize\sffamily\bfseries{#1}}}
   \newcommand{\margin}[1]{\marginpar{\scriptsize\sffamily\bfseries{#1}}}
\else
   \newcommand{\comment}[1]{}
   \newcommand{\margin}[1]{}
   
\fi
\ifx\ShowLineNumberVersion\AnswerYes
   \usepackage{lineno}
   \linenumbers
\fi
\ifx\ShowChangesVersion\AnswerYes
   \usepackage{ulem}
   \newcommand{\delete}[1]{\sout{#1}}            % delete #1 (strike-out)
   \renewcommand{\emph}[1]{\textit{#1}}           % ulem overwrites def of \emph as
\else

   \newcommand{\sout}[1]{}
   \newcommand{\xout}[1]{}

   \newcommand{\delete}[1]{}
\fi

\newcommand{\eg}{\textit{e.g.}\xspace}

\newcommand{\ie}{\textit{i.e.}\xspace}

\newcommand{\dis}{\displaystyle}

\newcommand{\hq}{\hspace{0.5em}}

\newcommand{\ii}{\mathrm{i}}

\newcommand{\vectorwithspace}[1]{\vec{#1}\mkern2mu\vphantom{#1}}
\newcommand{\vect}[1]{\vectorwithspace{#1}}

\newcommand{\ev}{\vectorwithspace{e}}

\newcommand{\qv}{\vectorwithspace{q}}

\newcommand{\kcotdelta}{\ensuremath{\mathrm{k}\!\cot\!\delta}}
\newcommand{\cotdelta}{\ensuremath{\cot\!\delta}}

\newcommand{\mpi}{\ensuremath{m_\pi}}

\newcommand{\fpi}{\ensuremath{f_\pi}}
\newcommand{\gA}{\ensuremath{g_A}}
\newcommand{\MeV}{\ensuremath{\mathrm{MeV}}}
\newcommand{\fm}{\ensuremath{\mathrm{fm}}}

\newcommand{\ChiEFT}{\foreignlanguage{greek}{q}EFT\xspace}

\newcommand{\NoPion}{\foreignlanguage{greek}{p}\hspace*{-0.38em}/\hspace*{0.1em}}
\newcommand{\EFTNoPion}{EFT(\NoPion)\xspace}

\newcommand{\LambdaNN}{\ensuremath{\overline{\Lambda}_{\N\N}}}

\newcommand{\OPE}{OPE\xspace}

\newcommand{\NXLO}[1]{N\ensuremath{{}^{#1}}LO\xspace}

\newcommand{\wave}[3]{\ensuremath{{}^{#1}\mathrm{#2}_{#3}}\xspace}
\newcommand{\oneS}{\wave{1}{S}{0}}

\newcommand{\threeS}{\wave{3}{S}{1}}
\newcommand{\threeD}{\wave{3}{D}{1}}
\newcommand{\threeSD}{\wave{3}{SD}{1}}

\newcommand{\N}{\ensuremath{\mathrm{N}}}

\newcommand{\SU}{\ensuremath{\mathrm{SU}}}

\newcommand{\MN}{\ensuremath{M}}%_\mathrm{N}}} % nucleon mass
\newcommand{\calO}{\ensuremath{\mathcal{O}}}

\newcommand{\refdoi}[1]{\doi{#1}}
\title{The Unitarity-Limit Expansion for Two Nucleons with Perturbative Pions:
  Digest and Ideas} \ShortTitle{Unitarity with Perturbative Pions: Digest
  and Ideas}

\author*[a]{Harald W.\ Grie\3hammer\orcidlink{0000-0002-9953-6512}}
\affiliation[a]{Institute for Nuclear Studies, Department of Physics, \\The
    George Washington University, Washington DC 20052, USA}

\emailAdd{hgrie@gwu.edu}
\abstract{Theorists love nontrivial fixed points. In the Unitarity Limit, the
  $\N\N$ $S$-wave binding energies are zero, the scattering lengths infinite,
  Physics is universal, \ie~insensitive to details of the interactions, and
  observables display richer symmetries, namely invariance under both scaling
  and Wigner's combined $\SU(4)$ transformation of spin and isospin.  In
  ``Pionless'' EFT, both are explicitly but weakly broken and hence
  perturbative in the Unitarity Window (phase shifts
  $45^\circ\lesssim\delta(k)\lesssim135^\circ$, \ie~momenta
  $k\approx\mpi$). This Unitarity Expansion provides strong hints that Nuclear
  Physics resides indeed in a sweet spot: bound weakly enough to be
  insensitive to the details of the nuclear interaction; and therefore
  interacting strongly enough that the $\N\N$ scattering lengths are
  perturbatively close to the Unitarity Limit. In this paradigm change,
  $\N\N$ details are less important than $\N\N\N$ interactions
  to explain the complexity and patterns of the nuclear chart.
  \\[1ex]
  This presentation is a digest of the first quantitative exploration of
  corrections to this picture when pions are included~\cite{Teng:2024exc} (see
  there for a more comprehensive list of references).
  Since the pion mass and decay constant introduce dimensionful scales in the
  $\N\N$ system, they explicitly break the symmetries of the Unitarity fixed
  point. In \ChiEFT, these symmetries must therefore be hidden and instead be
  classified as emergent.
  This text focuses on the \ChiEFT variant with Perturbative (``KSW'')
  Pions at next-to-next-to leading order (\NXLO{2}).
  In the \oneS channel up to cm momenta $\lesssim300\;\MeV$, the results are
  clearly converged order-by-order and agree very well with phase shift
  analyses. 
  Apparent large discrepancies in the \threeS channel even at
  $k\approx100\;\MeV$ are remedied by taking only the central part of the
  pion's \NXLO{2} contribution. In contradistinction to the tensor part, it
  does not mix the different Wigner-$\SU(4)$ multiplets and hence is identical
  in \oneS and \threeS.
  With this formulation, pionic effects are small in both channels even for
  $k\gtrsim\mpi$, \ie~where both Unitarity and pion effects are expected to be
  relevant.
  This leads to the \emph{Hypothesis} that both scale invariance and
  Wigner-$\SU(4)$ symmetry in the Unitarity Expansion show \emph{persistence},
  \ie~the footprint of both combined dominates even for $k\gtrsim\mpi$ and is
  more relevant than chiral symmetry, so that the tensor/Wigner-$\SU(4)$
  symmetry-breaking part of \OPE is suppressed and does not enter before
  \NXLO{3}.
  Included are also ideas about underlying mechanisms and LO results of
  \ChiEFT with Nonperturbative Pions in the expansion about Unitarity.}

\FullConference{The 11th International Workshop on Chiral Dynamics (CD2024)\\
 26-30 August 2024\\
Ruhr University Bochum, Germany\\}

\begin{document}
\maketitle

\noindent
\hfill\parbox{0.92\linewidth}{\emph{This contribution remembers Thomas
    C.~Mehen, who died unexpectedly a few months after this conference, in the
    waning days of 2024. His postdoctoral work on $\N\N$ systems with
    perturbative pions~\cite{Fleming:1999bs, Fleming:1999ee} and with
    Wigner-$\SU(4)$ symmetry~\cite{Mehen:1999qs} are fundamental to this
    discussion, and his work on relations to entanglement~\cite{Low:2021ufv,
      Liu:2022grf} is apt to provide further clues.}}

%%%%%%%%%%%%%%%%%%%%%%%%%%%%%%%%%%%%%%%%%%%%%%%%%%%%%%%%%%%%%%%%%%%%%%%%%%%%%
\section{Introduction}

Effective Field Theories (EFTs) of Nuclear Physics do not offer an explanation
why the observables of few-nucleon systems are dominated by anomalous scales:
The \oneS and \threeS $\N\N$ channels are very weakly bound, with associated
momentum scales of $-8\;\MeV$ and $+45\;\MeV$, respectively, set by the
inverse scattering lengths. These scales are much smaller than the pion mass
$\mpi\approx140\;\MeV$, the natural low-momentum QCD scale of Nuclear
Physics. Rather than explain, EFTs simply impose an ordering scheme whose
leading-order (LO) is found by iterating a $\N\N$ interaction infinitely often
to accommodate such anomalously shallow virtual and real bound states.
Perfectly valid and consistent versions of \ChiEFT exist in which LO is
perturbative and the binding energy of light nuclei is set by the scale
$\frac{\mpi^2}{\MN}$, with $\MN$ the nucleon mass. But these are not realised
in Nature.  A preference for highly-symmetric states could hold the key for
understanding
this. % Since the intrinsic momentum scales are so small, one may recover them
% by expanding about scale zero, namely about the Unitarity Limit.

So, consider the two-nucleon scattering amplitude at relative momentum
$k$ in the cm frame,
\begin{equation}
    \label{eq:amplitude}
    A(k)=\frac{4\pi}{M}\;\frac{1}{\kcotdelta(k)-\ii k}\;\;,
\end{equation}
where ``$-\ii k$'' ensures Unitarity, \ie~probability conservation, while
$\kcotdelta(k)$ parametrises the part which encodes all information on the
interactions. Depending on their relative sizes, any self-respecting theorist
is tempted to set up two fundamentally different expansions

\textbf{Born Approximation in the Born Corridor:} The phase shift is
``small''$|\delta(k)|\lesssim45^\circ$, \ie~$|\kcotdelta| \gtrsim |\ii k|$, so
that contributions from interactions are small and treated in perturbation,
\begin{equation}
  A(k)\Big|_\mathrm{Born}=\frac{4\pi}{M}\;\frac{1}{\kcotdelta(k)}
  \left[1+\frac{\ii}{\cotdelta(k)}+\calO(\cot^{-2}\delta)\right]\;\;,
\end{equation}
with the leading piece just given by the potential $V$:
$\dis\frac{1}{\kcotdelta(k)}\propto\langle\mathrm{out}|V|\mathrm{in}\rangle$. In
this expansion, interaction details are crucial to capture phase shifts
well. The amplitude has no pole (bound state).

\textbf{Unitarity Expansion in the Unitarity Window:} This is the focus of
this presentation. The
phase shift is ``large'', $45^\circ\lesssim|\delta(k)|\lesssim135^\circ$,
\ie~$|\kcotdelta|\lesssim|\ii k|$ about the Unitarity Point $\cotdelta=0$:
\begin{equation}
  \label{eq:Unitarity-amp}
  A(k)\Big|_\mathrm{Uni}=\frac{4\pi}{M}\;\frac{1}{-\ii k}
  \left[1+\frac{\cotdelta(k)}{\ii}
    +\frac{\cot^2\delta(k)}{\ii^2}+\calO(\cot^3\delta)\right]\;\;.
\end{equation}
In the Effective-Range Expansion,
$\kcotdelta=-\frac{1}{a}+\frac{r}{2}k^2+\dots$ with scattering length $a$ and
effective range $r$, this converges for an expansion parameter
$Q\sim\frac{1}{ak}\sim\frac{rk}{2}\ll1$.
Now, contributions from interactions are so strong that their details do not
matter as much as that they are simply very strong -- so strong indeed that
probability conservation limits their impact on observables and Unitarity
dominates.
In the language of Information Theory, it is critical that $a$ is large, but
its value is less important and rather on par with other subdominant
information (like from $r$).
An anomalously shallow bound state emerges naturally at LO since the
amplitude's pole is at zero. 

Since the Unitarity Limit has no intrinsic scale at LO, all dimensionless
$\N\N$ observables are zero or infinite, while all dimensionful quantities
(like cross sections) are homogeneous functions of $k$ whose dimensionless
coefficients do not depend on the interaction. Since details of the
interactions are washed out and can be treated in perturbation, differences
between systems at Unitarity can then only come from different symmetry
properties of the Unitarity Point itself but are universal
otherwise. Therefore, Unitarity and Universality are closely related. This
Unitarity Expansion has been proposed as key to the emergence of simple,
unifying patterns in complex systems like the nuclear chart; see
\eg~\cite{Konig:2016utl} and~\cite{vanKolck:2020plz, Kievsky:2021ghz} for
reviews as well as~\cite{Brown}.

\begin{figure}[!htbp]
    \begin{center}
      \includegraphics[width=0.97\linewidth]
      {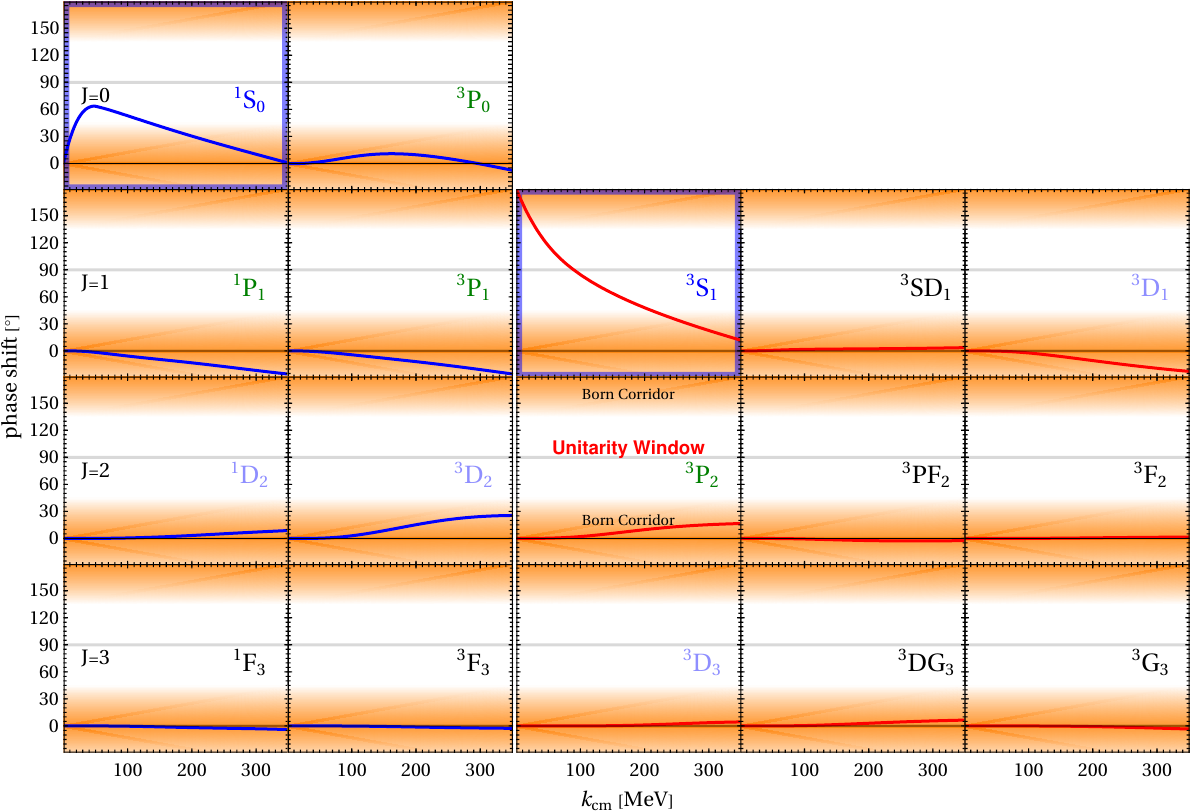}
  \caption{(Colour on-line) Born Corridor (shaded) and Unitarity Window
    (white) of $\N\N$  phase shifts for the $J\le3$ channels in the Nijmegen
    PWA~\cite{Stoks:1993tb} (Stapp-Ypsilanti-Metropolis
    (SYM/''bar'') parametrisation).}
    \label{fig:Unitaritywindow}
  \end{center}
  \vspace*{-3ex}
\end{figure}

How relevant is the Unitarity Window in the $\N\N$ system?
Figure~\ref{fig:Unitaritywindow} shows that only for two channels are phase
shifts clearly inside it, and only at momenta
$30\;\MeV\lesssim k\lesssim300\;\MeV$ (lab energies
$2\;\MeV\lesssim E_\mathrm{lab}\lesssim200\;\MeV$) which are relevant for
low-energy properties of nuclear systems: \oneS and \threeS (framed in
blue). Therefore, the Unitarity Expansion is \emph{only applicable} in these
two channels, and in none of the higher partial waves. Thus, this
presentation does not concern itself with the other partial waves -- not
because it wants to avoid describing them, but rather because the Unitarity
Expansion is simply inapplicable there, meaning the present investigation into
two-nucleon Unitarity is agnostic about them. Other systems can be a different
matter.

%%%%%%%%%%%%%%%%%%%%%%%%%%%%%%%%%%%%%%
%%%%%%%%%%%%%%%%%%%%%%%%%%%%%%%%%%%%%%

%%%%%%%%%%%%%%%%%%%%%%%%%%%%%%%%%%%%%%%%%%%%%%%%%%%%%%%%%%%%%%%%%%%%%%%%%%%%%
\section{The Unitarity Expansion with Perturbative Pions}

\begin{figure}[!b]
\begin{center}
     \includegraphics[width=\linewidth]{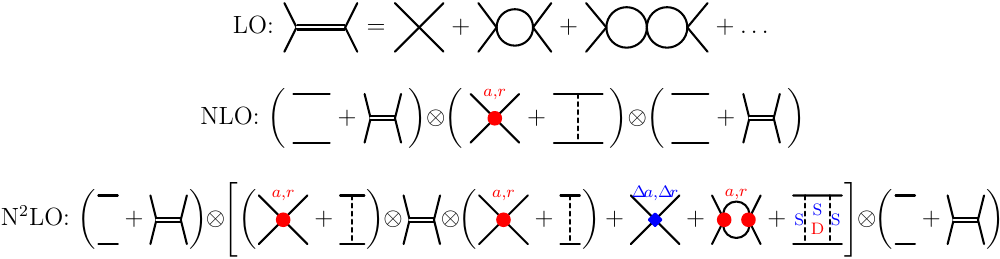}
     \caption{(Colour on-line) \ChiEFT with Perturbative Pions in the
       Unitarity Expansion at LO (top); NLO (middle) with CTs (red circle)
       fixed to scattering length $a$ and effective range $r$; \NXLO{2} with
       CTs (blue diamonds) so that $a$ and $r$ do not change from the
       respective NLO values. The last term in square brackets at \NXLO{2} is
       once-iterated \OPE, with intermediate orbital angular momentum as
       indicated.}
\label{fig:amplitudes}
\end{center}

\vspace*{-4ex}
\end{figure}

How is the Unitarity Window accommodated in \ChiEFT?  Let us investigate the
transition between ``pionless'' and ``pionic'' EFT by employing \ChiEFT ``with
Perturbative/KSW Pions'', proposed by Kaplan, Savage and
Wise~\cite{Kaplan:1998tg, Kaplan:1998we}. This is the only \ChiEFT which is
generally accepted to be self-consistent and renormalisable order by order,
with a well-understood power counting~\cite{Beane:2001bc}. Its dimensionless
expansion parameter $Q$ is
\begin{equation}
  \label{eq:Q}
  Q=\frac{k,\mpi}{\LambdaNN}\;\;\mbox{, with }
  \LambdaNN=\frac{16\pi\fpi^2}{\gA^2M}\approx300\;\MeV\;\;
\end{equation}
the scale at which iterations of one-pion exchange (\OPE) are not suppressed
any more. Its contributions up to and including \NXLO{2} are summarised in
fig.~\ref{fig:amplitudes}. Traditionally, the scattering lengths $a$ are set
to the physical values already at LO. The analytic results of such amplitudes
were derived by Rupak and Shoresh in the \oneS channel~\cite{Rupak:1999aa},
and by Fleming, Mehen and Stewart (FMS) in \threeSD~\cite{Fleming:1999bs,
  Fleming:1999ee}. The Unitarity Expansion simply demotes the contribution
from the scattering lengths, so that LO contains no scale but the scattering
momentum and knows only $\mathrm{S}$-wave interactions; NLO consists of
contributions from the scattering length, effective range $r$ and non-iterated
\OPE; and \NXLO{2} adds once-iterated \OPE, plus parameters which cancel its
contribution to $a$ and $r$, as these are already reproduced at NLO. It thus
adds an expansion parameter which can for practical purposes be taken to
be numerically similar to that of eq.~\eqref{eq:Q}: 
\begin{equation}
  \label{eq:Q2}
  Q=\frac{1}{(k,\mpi)a}\sim\frac{k,\mpi}{\LambdaNN}\ll1\;\;\mbox{ in \ChiEFT with
    KSW Pions about Unitarity.}
\end{equation}
The Unitarity Expansion becomes hence inapplicable both as $k\searrow0$ and
$k\nearrow\LambdaNN$.  This counting leads to quite a few
simplifications. Analytic amplitudes are given in ref.~\cite{Teng:2024exc},
plus further details on their analytic structure, on parameter choices, on
Bayesian and on-Bayesian estimates of theory uncertainties and on the
extraction of phase shifts from amplitudes.  The following compares results to
both the phenomenological phase shifts and to the Unitarity Expansion of the
EFT ``Without Pions'' (\EFTNoPion) which in this case reduces to the
effective-range expansion about $\frac{1}{a}=0$.

%%%%%%%%%%%%%%%%%%%%%%%%%%%%%%%%%%%%%%
%%%%%%%%%%%%%%%%%%%%%%%%%%%%%%%%%%%%%%

\begin{figure}[!t]
\begin{center}
  \includegraphics[width=0.49\textwidth]%[width=0.5\linewidth]
  {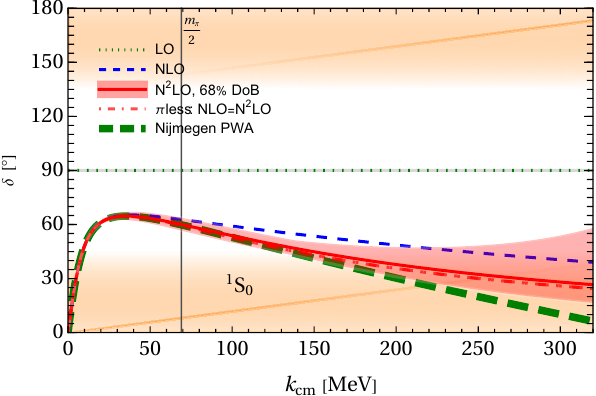}
  \caption{(Colour on-line) \oneS
    phase shift compared to the Nijmegen
    PWA~\cite{Stoks:1993tb} (thick green dashed). Green dotted: LO; blue
    dashed: NLO; red solid: \NXLO{2} with $68\%$ degree-of-belief (DoB)
    interval (Bayesian truncation uncertainty); red dash-dotted:
    NLO$=$\NXLO{2} in \EFTNoPion. Shaded: ``Born Corridors''
    of fig.~\ref{fig:Unitaritywindow}.}
\label{fig:results1S0}
\end{center}

\vspace*{-3ex}
\end{figure}

The \oneS phase shift, fig.~\ref{fig:results1S0}, converges well
order-by-order even as $k\to\LambdaNN$. Even well outside the Unitarity
Window, \NXLO{2} and PWA differ only as much as NLO and \NXLO{2} -- and still
less than LO and NLO. That indicates good order-by-order
convergence. Remarkably, explicit pionic degrees of freedom appear at
$k\gtrsim300\;\MeV $ to have a minuscule impact since \NXLO{2} is nearly
indistinguishable from the \NXLO{2} \EFTNoPion result. In this channel, we
have a self-consistent EFT with pions in all of the Unitarity Window, with a
breakdown scale of about $300\;\MeV\approx\LambdaNN$.

%%%%%%%%%%%%%%%%%%%%%%%%%%%%%%%%%%%%%%
%%%%%%%%%%%%%%%%%%%%%%%%%%%%%%%%%%%%%%

\begin{figure}[!h]
\begin{center}
  \includegraphics[width=0.49\textwidth]%[width=0.5\linewidth]
  {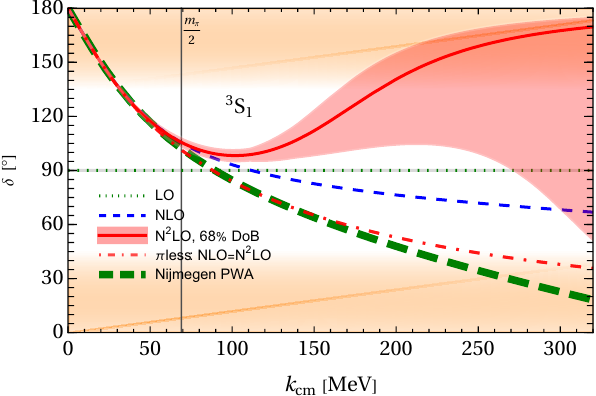}
  %\\
  \includegraphics[width=0.49\textwidth]%[width=0.5\linewidth]
  {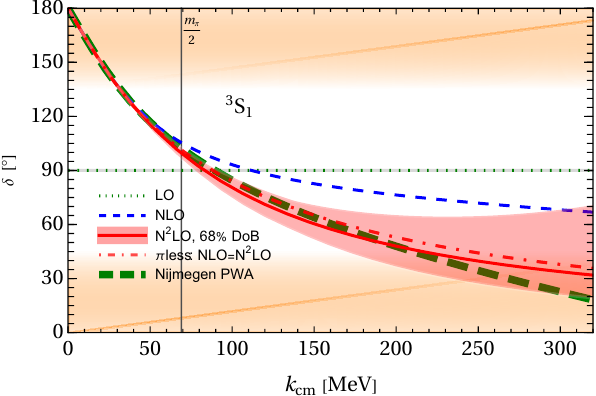}
  \caption{(Colour on-line) Phase shift in the \threeS channel compared to
    the Nijmegen PWA~\cite{Stoks:1993tb}.  Left: full \NXLO{2} amplitude.
    Right: Wigner-$\SU(4)$ symmetric part only. Details as in
    fig.~\ref{fig:results1S0}.}
\label{fig:results3S1}
\end{center}
\end{figure}

In contradistinction, the result for the \threeS channel is catastrophic, as
FMS already noticed; see left of fig.~\ref{fig:results3S1}. While NLO looks
reasonable by eye, \NXLO{2} deviates dramatically from the PWA just above the
\OPE branch-point scale, $\frac{\mpi}{2}$. This is the more puzzling as
\EFTNoPion agrees well with the PWA even at $k\gtrsim100\;\MeV$.  With pions,
the deviation between NLO and \NXLO{2} becomes around $k\approx\mpi$ as large
as the difference between LO and NLO, and larger than the deviation from the
PWA. The breakdown is hardly gradual but sudden, with no hint at NLO of the
unnaturally large \NXLO{2} curvature around $100\;\MeV$. This is the more
concerning as phase shifts are there well inside the Unitarity Window. Pions
at \NXLO{2} appear to have an outsized and wrong impact for
$k\gtrsim100\;\MeV$, and most of the Unitarity Window lies outside the radius
of convergence. That is unsatisfactory.

%%%%%%%%%%%%%%%%%%%%%%%%%%%%%%%%%%%%%%
%%%%%%%%%%%%%%%%%%%%%%%%%%%%%%%%%%%%%%

What is the origin of the stark discrepancy between the \oneS and \threeS
channels in both convergence and reasonable range of applicability? Recall
that LO is independent of the scattering length, \ie~both amplitudes are
identical and hence invariant under Wigner's combined-$\SU(4)$ spin and
isospin rotations~\cite{Wigner, Hund, Mehen:1999qs}. With scale invariance,
this symmetry emerges therefore naturally in the Unitarity Limit. The
long-range part of \OPE, however, appears to break both:
\begin{equation}
  \label{eq:OPE}
  \begin{split}
  V_\mathrm{OPE}&=-\frac{\gA^2}{12\fpi^2}\;\frac{\qv^2}{\qv^2+\mpi^2}
  \bigg[
    \left(\vect{\sigma}_1\cdot\vect{\sigma}_2\right)+
  [3\left(\vect{\sigma}_1\cdot\ev_q\right)
    \left(\vect{\sigma}_2\cdot\ev_q\right)
    -\left(\vect{\sigma}_1\cdot\vect{\sigma}_2\right)]\bigg]
  \left(\vect{\tau}_1\cdot\vect{\tau}_2\right)\\[0.5ex]
  &=:\left(\vect{\sigma}_1\cdot\vect{\sigma}_2\right)
    \left(\vect{\tau}_1\cdot\vect{\tau}_2\right) V_C+
    [3\left(\vect{\sigma}_1\cdot\ev_q\right)
    \left(\vect{\sigma}_2\cdot\ev_q\right)
    -\left(\vect{\sigma}_1\cdot\vect{\sigma}_2\right)]
  \left(\vect{\tau}_1\cdot\vect{\tau}_2\right) V_T\;\;.
  \end{split}
\end{equation}
As $\fpi$ and $\mpi$ carry mass dimensions, they explicitly break scale
invariance. On top of that, the spin-isospin structure of the tensor part,
$V_T$, induces $\mathrm{S}\leftrightarrow\mathrm{D}$ and
$\mathrm{D}\to\mathrm{D}$ transitions. These are only possible in \threeS, and
not in \oneS where the tensor piece is of course identically zero. Such
partial-wave mixing therefore lifts the Wigner-$\SU(4)$ symmetry between the
$\mathrm{S}$ waves -- in a possibly slight abuse of language, this can be
called ``Wigner-$\SU(4)$ symmetry breaking''.  On the other hand, the central
part, $V_C$, is manifestly Wigner-$\SU(4)$ symmetric, \ie~it contributes only
in $\mathrm{S}\to\mathrm{S}$ transitions and is hence identical in the \oneS and
\threeS channels. % It is also the \emph{only}
% contribution in the \oneS channel. 
Therefore, only amplitudes without $V_T$ are Wigner-invariant, while those
with at least one $V_T$ automatically break the symmetry. Consequently, it is
not self-understood how \ChiEFT's explicit pionic degrees of freedom can be
reconciled with the symmetries of the Unitarity Expansion which they break
rather strongly, and how one can thus extend the Unitarity Expansion in
\ChiEFT to the whole Unitarity Window, including $k\gtrsim\mpi$.

With perturbative pions, no Wigner-$\SU(4)$ symmetry-breaking \OPE
contribution enters at NLO, simply because a single \OPE is sandwiched between
the LO pure-$\mathrm{S}$ wave amplitudes (which are Wigner-$\SU(4)$
symmetric), disallowing $\mathrm{S}\leftrightarrow\mathrm{D}$ mixing. The only
Wigner-$\SU(4)$ breaking terms at NLO come from scattering lengths and
effective ranges. They are explicit but weak and, as \EFTNoPion shows, can
well be captured in perturbation about the Unitarity Limit $\frac{1}{a}=0$.

At \NXLO{2}, however, sandwiching once-iterated \OPE between the LO \threeS
waves does allow for an $\threeS\to\threeD\to\threeS$ transition,
\ie~$\mathrm{D}$-wave propagation between each \OPE (last term in square
brackets of the bottom of fig.~\ref{fig:amplitudes}). This piece is of course
absent in the \oneS channel. \OPE breaks thus Wigner-$\SU(4)$ symmetry in
$\mathrm{S}$-waves only when it is once-iterated, namely at \NXLO{2}, but not
earlier.

This led to the idea to simply \emph{impose} Wigner-$\SU(4)$ symmetry on the
pion at \NXLO{2}: eliminate the $V_T$ part! The qualitative and
quantitative improvement is obvious; see left graph of
fig.~\ref{fig:results3S1}. Similar to \oneS, the phase shift converges now
order-by-order even as $k\to\LambdaNN$ just outside the
Unitarity Window. The difference of \NXLO{2} and PWA is wholly within the
Bayesian $68\%$ DoB band, even smaller than in the \oneS channel and quite a
bit smaller than the shift from NLO to \NXLO{2}. With only small differences
to \EFTNoPion, pionic degrees of freedom have again a minuscule impact even at
$k\gtrsim\mpi$. The empirical breakdown scale is around
$k\gtrsim250\;\MeV\approx\LambdaNN$, as expected.

Thus, a Wigner-$\SU(4)$ invariant \ChiEFT appears well-suited to describe
perturbative-pionic Physics in both $\mathrm{S}$ waves up to a common
breakdown scale of at least $k\gtrsim250\;\MeV\approx\LambdaNN$, plus good
agreement with PWAs even at these high momenta. The (non-analytic parts of)
pionic effects appear then very small, as comparison to \EFTNoPion shows.

Reference~\cite{Teng:2024exc} includes detailed discussions of Bayesian
order-by-order convergence, of other mutually consistent semi-quantitative
theory uncertainty estimates, and of empirical determinations of the breakdown
scale. All confirm the outline above.

\begin{figure}[!t]
\begin{center}
  \includegraphics[width=0.483\textwidth]%[width=0.5\linewidth]
  {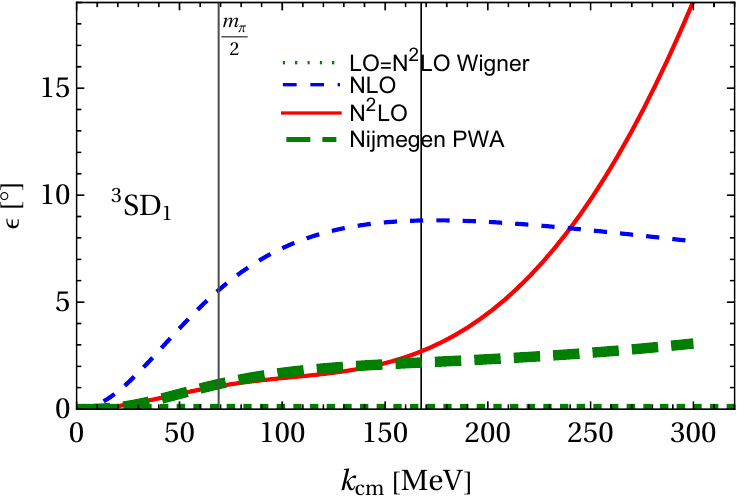}\hfill%
  \includegraphics[width=0.49\textwidth]%[width=0.5\linewidth]
  {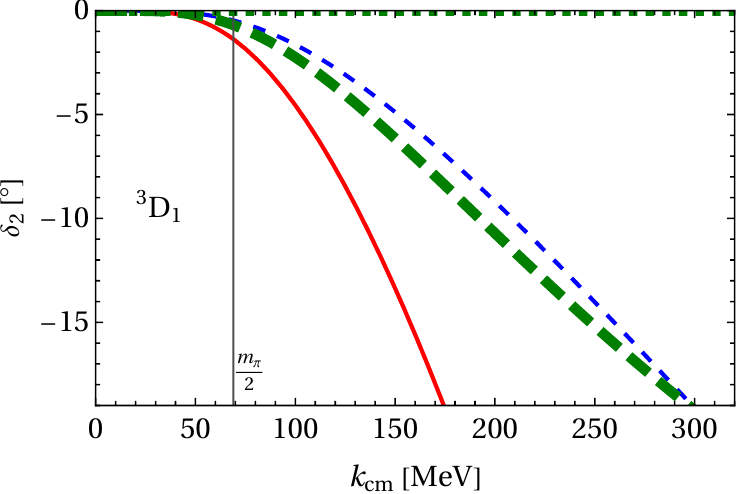}
  \caption{(Colour on-line) Mixing angle (left) and \threeD phase shift
    (right) in the \threeSD channel, compared to the Nijmegen
    PWA~\cite{Stoks:1993tb} (thick green dashed). Details as in
    fig.~\ref{fig:results1S0}.}
\label{fig:results3SD1}
\end{center}

\vspace*{-3ex}
\end{figure}

Such an approach solves another puzzle described by FMS. As shown in
fig.~\ref{fig:results3SD1}, both $\wave{3}{SD}{1}$ mixing and \threeD phase
shift are terrible with the full \NXLO{2} amplitudes. The mixing angle may
seem to agree well with the PWA, but the \NXLO{2} correction is nearly as
large as NLO itself for $k\gtrsim50\;\MeV$, \ie~order-by-order convergence is
extremely poor. It is even worse in \threeD. On the other hand, imposing
Wigner-$\SU(4)$ symmetry eliminates $V_T$, so that the partial waves do not
mix at all at \NXLO{2}. Assuming that it only enters at \NXLO{3} or higher,
one can estimate the typical magnitude of phase shift and mixing angle at,
say, $k\approx\mpi$ as $Q^3\approx\left(\frac{\mpi}{\LambdaNN}\right)^3$ of
the LO phase shift $\delta_\mathrm{LO}=90^\circ$, namely about
$10^\circ$. That is not inconsistent with the PWA values. Remember also that
further contact interactions enter at \NXLO{3} which contribute to
$\mathrm{SD}$ mixing and $\mathrm{DD}$ transitions. These may help reproduce
more-accurate partial wave mixing.

%%%%%%%%%%%%%%%%%%%%%%%%%%%%%%%%%%%%%%
%%%%%%%%%%%%%%%%%%%%%%%%%%%%%%%%%%%%%%

%%%%%%%%%%%%%%%%%%%%%%%%%%%%%%%%%%%%%%%%%%%%%%%%%%%%%%%%%%%%%%%%%%%%%%%%%%%%%
\section{Ideas}

This evidence is not inconsistent with a \textbf{Hypothesis}: The symmetries
of the Unitarity Limit are broken weakly in Nuclear Physics. They show
\emph{persistence}, \ie~the footprint of both combined in observables at
$k\gtrsim\mpi$ is more relevant than chiral symmetry. In particular, the
tensor/Wigner-$\SU(4)$ symmetry-breaking part of one-pion exchange in the
$\N\N$ \threeS channel of \ChiEFT with Perturbative (KSW) Pions is
super-perturbative, \ie~suppressed and does not enter before \NXLO{3}.

\begin{figure}[!t]

\vspace*{-3ex}
\begin{center}
  \includegraphics[width=0.46\textwidth]%[width=0.5\linewidth]
  {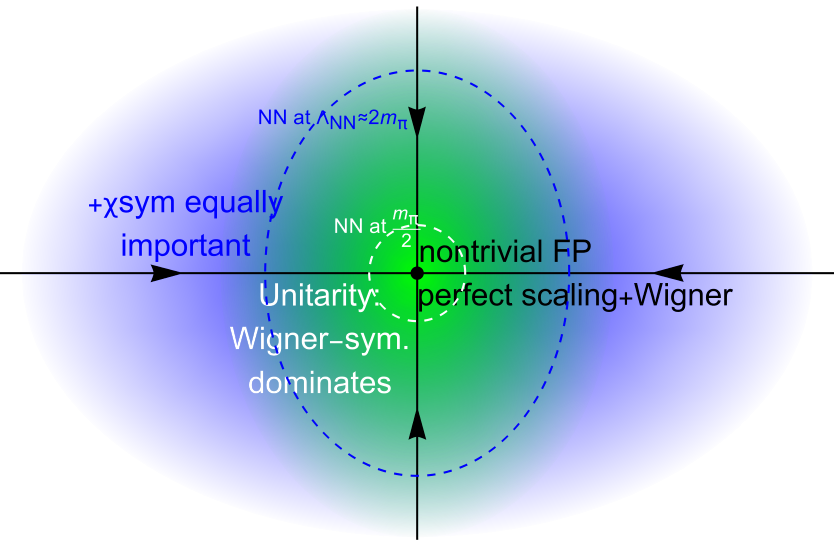}
  \caption{(Colour on-line) Sketch of an idea of symmetries around the
    Unitarity fixed point.}
\label{fig:fixedpoint}
\end{center}

\vspace*{-3ex}
\end{figure}

If so, then the picture of fig.~\ref{fig:fixedpoint} emerges around the
renormalisation-group fixed point (FP) of Unitarity in two-nucleon systems:
The FP is of course non-Gau\3ian (LO is nonperturbative), and its universality
class is all theories with Wigner-$\SU(4)$ symmetry (and scale invariance
since one is at a FP). In its immediate vicinity, the weak breaking of scaling
and Wigner-$\SU(4)$ symmetry dominate, while chiral symmetry is
subdominant. \EFTNoPion is the EFT of \ChiEFT at low momenta but has no
explicit chiral symmetry. If chiral symmetry were thus important right around
the FP, then \ChiEFT would lie in a different universality class than
\EFTNoPion. In this scenario, the Unitarity FP protects Wigner-$\SU(4)$
symmetry to be only weakly broken, while chiral symmetry in the few-nucleon
sector has no such strong protection, simply because it is not a
characteristic symmetry of the FP. It \emph{can} therefore be broken
substantially, leading to suppression of the tensor part $V_T$. Further away
from the FP, chiral symmetry becomes as important as Wigner-$\SU(4)$ symmetry,
and will eventually dominate for large enough momenta -- as will the \OPE's
tensor piece. The scale $\LambdaNN$ at which \OPE becomes nonperturbative is
an obvious candidate for this inversion. Since the zero- and one-nucleon
sector of \ChiEFT are perturbative, \ie~the projection of the FP onto these is
Gau\3ian, this does not affect their chiral counting.

%%%%%%%%%%%%%%%%%%%%%%%%%%%%%%%%%%%%%%
%%%%%%%%%%%%%%%%%%%%%%%%%%%%%%%%%%%%%%

%%%%%%%%%%%%%%%%%%%%%%%%%%%%%%%%%%%%%%
\begin{figure}[!b]
\begin{center}
  \pbox{\linewidth}{\includegraphics[width=0.48\textwidth]
  {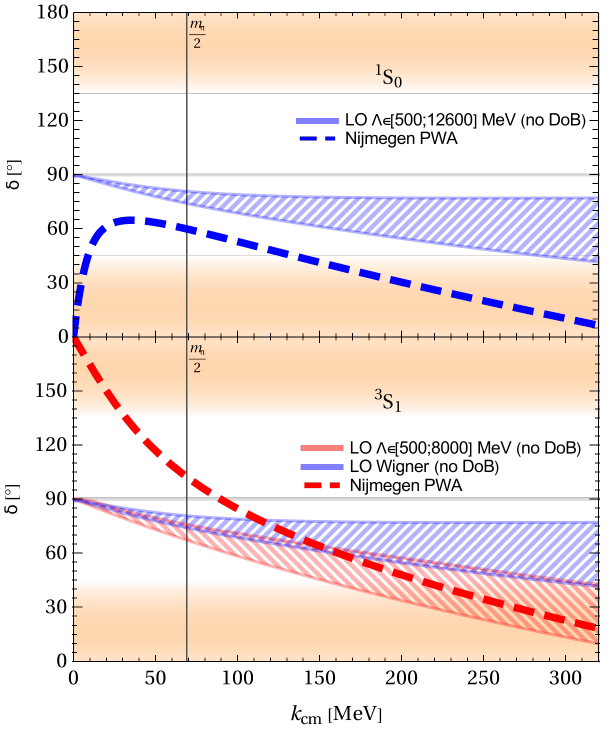}}\hq
  \pbox{\linewidth}{\includegraphics[width=0.37\textwidth]
  {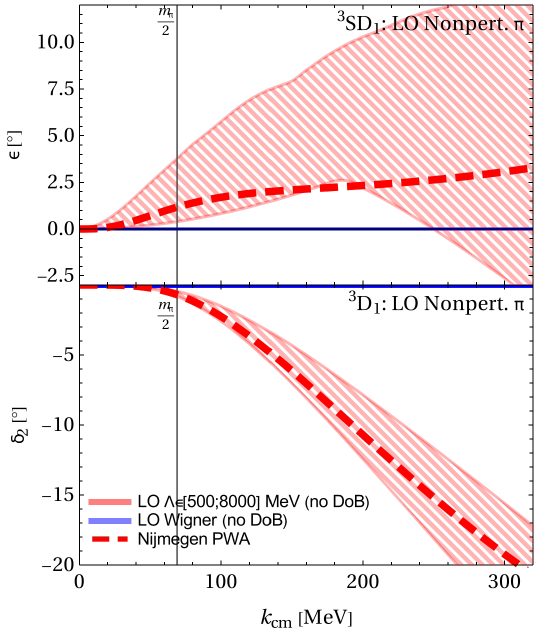}}
  \caption{(Colour on-line) Phase shifts with nonperturbative pions at LO in
    the Unitarity Limit ($\frac{1}{a}=0$), compared to the Nijmegen
    PWA~\cite{Stoks:1993tb}. Left: \oneS (top) and \threeS (bottom)
    channels. Right: \threeSD mixing angle (top right) and \threeD phase shift
    (bottom right), compared to the Nijmegen PWA~\cite{Stoks:1993tb}. Blue:
    Wigner-$\SU(4)$ symmetric part of \OPE. Red: full \OPE. The hatched areas are a
    rough estimate of higher-order corrections from varying the cutoff between
    $500$ and $12,600\;\MeV$, \emph{not} Bayesian degree-of-belief intervals.}
\label{fig:nonpertPions}
\end{center}

\vspace*{-4ex}
\end{figure}
%%%%%%%%%%%%%%%%%%%%%%%%%%%%%%%%%%%%%%

If this Hypothesis is to be more than a hunch, then a small, dimensionless
(systematic) expansion parameter rooted in Wigner-$\SU(4)$ symmetry must be
found to decide both \emph{a priori} and semi-quantitatively at which order
Wigner-$\SU(4)$ breaking pion contributions and correlated two-pion exchange
enter.
Ref.~\cite{Teng:2024exc} discusses two candidates, both somewhat problematic:
Large-$N_C$ and entanglement. It also contains a shopping list of processes in
which the Hypothesis can be tested, albeit the Goldstone mechanism converting
global chiral symmetry into local, weakly interacting field excitations may
eventually be too strong to overcome and lead to the Hypothesis'
downfall. What is called for here is unusual since we are more used to
promoting interactions in the chiral counting based on renormalisation-group
arguments, than demoting them as in the Hypothesis.

Clearly, one should also explore how Wigner-$\SU(4)$ symmetry emerges inside
the Unitarity Window in \ChiEFT with \emph{non-}perturbative pions. LO results
in fig.~\ref{fig:nonpertPions} indicate that the situation may not be
hopeless.  The \threeSD results are again decomposed into full and
Wigner-$\SU(4)$ symmetric pieces. The hatched corridors are not Bayesian
degree-of-belief intervals but come from simply varying the momentum cutoff,
so uncertainties are likely under-estimated. The one momentum-independent
counter term per channel is determined by the Unitarity Limit,
$\frac{1}{a}=0$. Since \OPE now enters already at LO, explicit scale breaking
can already be seen, albeit it appears to be small. The predicted effective
ranges are $[1\dots2]\;\fm$ in the \oneS channel and Wigner-$\SU(4)$ symmetric
case, and $[1.5\dots2.5]\;\fm$ in the full \threeSD case. The PWA value of
$2.77\;\fm$ for \oneS appears superficially less compatible with that than the
$1.85\;\fm$ for \threeSD. The \oneS phase needs further study; the \threeS
phase shift might show a slight preference for the full version over the
Wigner-$\SU(4)$ symmetric one. The full version is compatible with the
phenomenological $\mathrm{SD}$ mixing and \threeD phase, but especially the
former comes with humongous uncertainties.  More work is clearly needed,
especially on more reliable theory uncertainties, and a study of higher orders
is ongoing.

%%%%%%%%%%%%%%%%%%%%%%%%%%%%%%%%%%%%%%%%%%%%%%%%%%%%%%%%%%%%%%%%%%%%%%%%%%%%%
\section{Concluding Questions}

This contribution summarised and extemporated on a study in \ChiEFT with
Perturbative Pions~\cite{Teng:2024exc}. Universality (insensitivity of
amplitudes on details of interactions) and the associated Wigner-$\SU(4)$
symmetry emerge naturally in \EFTNoPion. Scattering lengths as fit parameters
are simply set to infinity, and the dominant interactions in the Lagrangean
are automatically scale and Wigner-$\SU(4)$ symmetric. But that is not
manifest in \ChiEFT. Rather, both are \emph{a priori} hidden and only
reconstructed as relevant by consulting data: Fitted to phase shifts, the
coefficient of the Wigner-$\SU(4)$ symmetric two-nucleon interaction
$C_S\;\mathbbm{1}$ is much larger than that of
$C_T\;\vec{\tau}_1\cdot\vec{\tau}_2$ which splits the \oneS-\threeS
Wigner-$\SU(4)$ multiplet. Therefore, both Wigner-$\SU(4)$ symmetry and
Universality/Unitarity become \emph{emergent} phenomena in \ChiEFT -- how is
not yet clear.

Can imposing a preference for Unitarity as a highly symmetric state about
which to expand provide a quantitative answer to fundamental questions: Why is
fine-tuning preferred? How does varying $\mpi$ affect the conclusions?
In-how-far is the Unitarity Limit compatible with the chiral limit? The
Hypothesis and study of ref.~\cite{Teng:2024exc} is merely a first proposal to
merge two highly successful concepts of Nuclear Theory, namely the expansion
about Unitarity and \ChiEFT, for mutual benefit and a better understanding of
why fine-tuning emerges in low-energy Nuclear Physics.

%%%%%%%%%%%%%%%%%%%%%%%%%%%%%%%%%%%%%%%%%%%%%%%%%%%%%%%%%%%%%%%%%%%%%%%%%%%%%
%%%%%%%%%%%%%%%%%%%%%%%%%%%%%%%%%%%%%%%%%%%%%%%%%%%%%%%%%%%%%%%%%%%%%%%%%%%%%
\subsection*{Acknowledgements}

To work on this project with MSc student Yu-Ping
Teng\orcidlink{0009-0007-8663-2998} was a pleasure.
Feedback was incorporated from this conference, from \textsc{Effective Field
  Theories and Ab Initio Calculations of Nuclei} (Nanjing, China), the ECT*
workshop \textsc{The Nuclear Interaction: Post-Modern Developments},
\textsc{INT-24-3 Quantum Few- and Many-Body Systems in Universal Regimes}, and
\textsc{INT-25-92W Chiral EFT: New Perspectives} (INT-PUB-24-052 and
INT-PUB-25-012). I am grateful to all organisers and participants for highly
stimulating atmospheres, and especially to S.~R.~Beane, U.~van Kolck, D.~Lee,
B.~Long, G.~A.~Miller, D.~R.~Phillips, M.~S\'anchez S\'anchez, R.~P.~Springer,
M.~J.~Savage and I.~Stewart for discussions and clarifications.
This work was supported in part by the US Department of Energy under contract
DE-SC0015393, and by George Washington University: by the Office of the Vice
President for Research and the Dean of the Columbian College of Arts and
Sciences; by an Enhanced Faculty Travel Award of the Columbian College of Arts
and Sciences. This research was conducted in part in GW's Campus in the
Closet.

%%%%%%%%%%%%%%%%%%%%%%%%%%%%%%%%%%%%%%%%%%%%%%%%%%%%%%%%%%%%%%%%%%%%%%%%%%%%%
%%%%%%%%%%%%%%%%%%%%%%%%%%%%%%%%%%%%%%%%%%%%%%%%%%%%%%%%%%%%%%%%%%%%%%%%%%%%%%

\end{document}